\pdfoutput=1

\documentclass[%
reprint,
superscriptaddress,
amsmath,amssymb,
aps,
prb,
]{revtex4-2}

\usepackage{graphicx}
\usepackage{color}

\usepackage{amssymb}
\usepackage{amsmath}
\usepackage{amsthm}
\usepackage{ascmac}
\usepackage{braket}
\usepackage[
    format=hang,
    justification=RaggedRight,
    font=small,
]{caption}
\usepackage{comment}
\usepackage{diagbox}
\usepackage{empheq}
\usepackage{hhline}
\usepackage{mathrsfs}
\usepackage{mathtools}
\mathtoolsset{showonlyrefs=false}
\usepackage{multirow}
\usepackage[version=3]{mhchem}
\usepackage{siunitx}
\usepackage[subrefformat=parens]{subcaption}
\usepackage{textgreek}
\usepackage{url}
\usepackage{upgreek}

\usepackage{dcolumn}
\usepackage{natbib}
\usepackage{textcase}
\usepackage{txfonts}
\usepackage{hyperref}
\hypersetup{
    colorlinks=true,
    linkcolor=blue,
    citecolor=blue,
    urlcolor=blue,
}

\newcommand{\mi}{\mathrm{i}}
\newcommand{\mpi}{\uppi}
\newcommand{\md}{\mathrm{d}}

\newcommand{\cre}[1]{\hat{#1}^{\dag}}

\allowdisplaybreaks[4]
\newcommand{\bsym}[1]{\boldsymbol{#1}}
\newcommand{\bothref}[2]{\ref{#1}\subref{#2}}

\makeatletter
\newcommand{\subscripts}[3]{%
    \@mathmeasure\z@\displaystyle{#2}%
    \global\setbox\@ne\vbox to\ht\z@{}\dp\@ne\dp\z@
    \setbox\tw@\box\@ne
    \@mathmeasure4\displaystyle{\copy\tw@_{#1}}%
    \@mathmeasure6\displaystyle{{#2}_{#3}}%
    \dimen@-\wd6 \advance\dimen@\wd4 \advance\dimen@\wd\z@
    \hbox to\dimen@{}\mathop{\kern-\dimen@\box4\box6}%
}
\makeatother

\begin{document}

\title{
Quantum Anomalous Hall Effect in Three-dimensional Topological Insulator/Thin-film Ferromagnetic Metal Bilayer Structure
}

\author{Katsuhiro Arimoto}
\email{k.arimoto@cmpt.phys.tohoku.ac.jp}
\affiliation{Department of Physics, Tohoku University, Sendai, Miyagi, 980-8578, Japan}
\affiliation{Institute for Materials Research, Tohoku University, Sendai, Miyagi, 980-8577, Japan}

\author{Takashi Koretsune}
\affiliation{Department of Physics, Tohoku University, Sendai, Miyagi, 980-8578, Japan}

\author{Kentaro Nomura}
\email{nomura@imr.tohoku.ac.jp}
\affiliation{Institute for Materials Research, Tohoku University, Sendai, Miyagi, 980-8577, Japan}
\affiliation{Center for Spintronics Research Network, Tohoku University, Sendai, Miyagi, 980-8577, Japan}

\begin{abstract}
We theoretically show that the three-dimensional (3D) topological insulator (TI)/thin-film ferromagnetic metal (FMM) bilayer structure is possible to be a quantum anomalous Hall (QAH) insulator with a wide global band gap. Studying the band structure and the weight distributions of eigenstates, we demonstrate that the attachment of a metallic thin-film on the 3DTI can shift the topologically non-trivial state into the metal layers due to the hybridization of bands around the original Dirac point. By introducing the magnetic exchange interaction in the thin-film layers, we compute the anomalous Hall conductivity and magnetic anisotropy of the bilayer structure to suggest the appearance of wider gap realizing QAH effect than usual materials, such as magnetically doped thin-films of 3DTI and 3DTI/ferromagnetic insulator heterostructures. Our results indicate that the 3DTI/thin-film FMM bilayer structure may implement the QAH effect even at room temperature, which will pave a way to the experimental realization of other exotic topological quantum phenomena.
\end{abstract}

\maketitle

\section{Introduction}
Three-dimensional (3D) topological insulators (TIs) \cite{Hasan-RMP82-3045,Qi-RMP83-1057} are materials which exhibit novel conductive state only on their surfaces. The surface state corresponds to linear-dispersing energy band with helical spin texture, called as Dirac surface state (DSS). The peculiar energy band connects topologically non-trivial bulk valence band and conduction band, which occurs due to a strong spin-orbit interaction in the 3DTI. The gapless DSS are robust against any perturbations unless time reversal symmetry in the 3DTI is broken or the bulk band gap is closed \cite{Shindou-PRB79-045321,Kobayashi-PRL110-236803}. Several actual materials, such as \ce{Bi_{$1-x$}Sb_{$x$}} alloy and \ce{Bi2Se3}-family tetradymite chalcogenides, have been experimentally proved to be 3DTI \cite{Hsieh-Nature452-970,Xia-NatPhys5-398,Sato-PRL105-136802,Kuroda-PRL105-146801,Chen-PRL105-266401}.

Quantum anomalous Hall (QAH) effect \cite{Nagaosa-RMP82-1539,Liu-AnnRevs7-301} is a two-dimensional phenomenon which gives discrete Hall conductivity in a unit of $e^{2}/h$ without external magnetic field. The quantized Hall conductivity directly reflects the Chern number of filled energy bands, which equals to the number of channel of chiral edge conduction when the bulk is insulating. This phenomenon typically requires strong spin-orbit interaction and ferromagnetic order to break time-reversal symmetry, which indicates that the TIs with ferromagnetism are candidate materials for QAH insulators \cite{Yu-Science329-61}. When the chemical potential lies in the gap of the Dirac dispersion induced by breaking of time-reversal symmetry, called as surface gap, the QAH effect occurs. Emergence of the energy gap indicates a possibility to realize other exotic topological phenomena, such as topological electromagnetic effect \cite{Essin-PRL102-146805,Li-NatPhys6-284,Wang-PRB93-045115,Sekine-arXiv2011-13601-v1} and image magnetic monopole effect \cite{Qi-Sciencs323-1184}, both of which were theoretically predicted from the topological field theory \cite{Qi-PRB78-195424}. The first observation of the QAH effect is in \ce{Cr}-doped thin-film \ce{(\mathrm{Bi,Sb})_{2}Te_{3}} \cite{Chang-Science340-167,Checkelsky-NatPhys10-731,Kou-PRL113-137201,Kou-SSC215-34}. In this case, the long range ferromagnetic order induced by the dopant opens the surface gap. However, the doping engenders spatial inhomogeneity of electronic potentials and exchange coupling by magnetic dopants. This inhomogeneity may cause spatial variation and diminishment of the surface gap \cite{Nomura-PRL106-166802,Lee-PNAS112-1316} and residual resistance by dissipative conductive states in bulk, both of which can make observable temperature of the QAH effect lower than the Curie temperature as much as two order of magnitude \cite{Chen-Science329-659}. For experimental confirmation of the exotic topological quantum phenomena, these disorder effects originating from the magnetic impurity should be solved and a wider surface gap in the whole Brillouin zone (BZ) needs to be induced.

Since the first observation of QAH effect in the thin-film 3DTI with magnetic dopants, there have been many related experiments with various geometries consisting of a 3DTI and ferromagnetic order. Mogi \textit{et al.} has reported that the surface gap can be enhanced by introducing \ce{Cr}-doped thin layers at the vicinity of both surfaces of a thin-film 3DTI due to suppression of the disorder on the surfaces \cite{Mogi-APL107-182401}. To realize the QAH insulator with the 3DTI bearing no disorder, 3DTI/ferromagnetic insulator heterostructures have been investigated \cite{Alegria-APL105-053512,Mogi-PRL123-016804}. In this structure, it had been anticipated that the strong exchange interaction acts on the DSS by penetration of the magnetization into the 3DTI surfaces with no disorder, resulting in a wider surface gap to the Dirac dispersion. However, the experimental results show non-quantized anomalous Hall conductivity, which indicates that the exchange interaction between the DSS and the penetrating magnetization is too small to open a considerable surface gap. Recently, Deng \textit{et al.} has observed the QAH effect in intrinsic magnetic TI, an odd-layer flake of \ce{MnBi2Te4} \cite{Deng-Science367-895}. Though \ce{MnBi2Te4} shows antiferromagnetism in 3D bulk \cite{Otrokov-Nature576-416,Swatek-PRB101-161109,Chen-PRX91-041040,Hao-PRX91-041038,Mong-PRB81-245209}, its odd-layer film cause the QAH effect. However, the observable temperature is one order lower than the N\'{e}el temperature. It still remains challenging to manage the wide surface gap for high-temperature QAH effect by operating significant exchange interaction to the DSS.

In this paper, we propose the possibility of high-temperature QAH effect arising in 3DTI/thin-film ferromagnetic metal (FMM) bilayer structure. This study is motivated by the experimental discovery of topological proximity effect; migration of the DSS into an attached thin-film metal \cite{Shoman-NatCommun6-6547}. From this discovery, we anticipate in the 3DTI/FMM bilayer structure that the DSS migrates into the FMM layer and is directly influenced by the strong exchange interaction. By calculating the energy band structure, the anomalous Hall conductivity, and the magnetic anisotropy, we will demonstrate this anticipation by showing the emergence of a wide QAH gap with no magnetic dopant in the 3DTI. The result suggests that the 3DTI/FMM bilayer structure can realize higher-temperature QAH effect and other exotic effects, which is anticipated to be applicable for low-power consumption devices and spintronics \cite{Tokura-NatRevPhys1-126}.

This paper is organized as follows. In Sec. \ref{sec:ModelHamiltonian}, we introduce a model Hamiltonian of a 3DTI/thin-film FMM bilayer structure shown in Fig. \ref{fig:System}. In Sec. \ref{sec:Result}, we analyze how the attachment of thin-film FMM acts the exchange interaction on the DSS in detail to clearly understand a mechanism of the appearance of the QAH gap in the bilayer structure. At first, in Sec. \ref{sec:TPE}, we discuss that the attachment of a thin-film normal metal (NM) on the 3DTI can shift the DSS on the 3DTI surface into the metal layer by computing the band structures and the layer-resolved weight distributions. Next, in Sec. \ref{sec:QAHE}, we demonstrate the QAH effect in the 3DTI/thin-film FMM bilayer structure. Then in Sec. \ref{sec:MagneticAnisotropy}, we calculate the magnetic-anisotropy energy to discuss the uniform magnetic configuration in this system from the viewpoint of band topology. We also compute variation of the QAH-gap size by changing the position of energy band of thin-film FMM in Sec. \ref{sec:Condition}. We discuss that the Dirac dispersion of the DSS and FMM band should initially possess energetical degeneracies near to the Dirac point to arise the wide QAH gap. Sec. \ref{sec:Conclusion} is devoted to conclusion.

\section{Model Hamiltonian} \label{sec:ModelHamiltonian}
\begin{figure}[tbhp]
	\centering
	\includegraphics[keepaspectratio, scale=0.5]{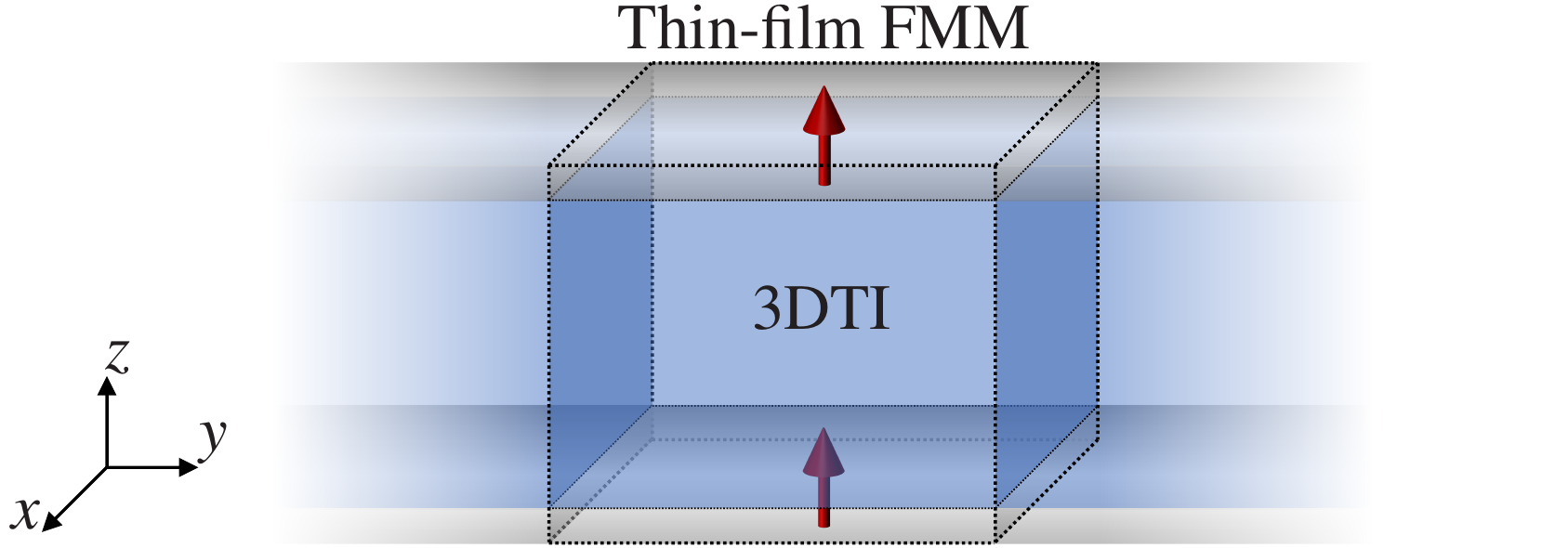}
	\caption{Schematic picture of the 3DTI/thin-film FMM bilayer. The red arrows represent magnetization vectors.}
	\label{fig:System}
\end{figure}
In this section, we construct a model Hamiltonian describing the structure shown in Fig. \ref{fig:System}. Our model Hamiltonian consists of three terms,
\begin{equation}
	\hat{\mathcal{H}}=\hat{\mathcal{H}}_{\mathrm{TI}}+\hat{\mathcal{H}}_{\mathrm{FMM}}+\hat{\mathcal{H}}_{\mathrm{IH}}, \label{eq:Hami_tot}
\end{equation}
where $\hat{\mathcal{H}}_{\mathrm{TI}}$, $\hat{\mathcal{H}}_{\mathrm{FMM}}$ and $\hat{\mathcal{H}}_{\mathrm{IH}}$ are 3DTI term, thin-film FMM term, and interface-hopping term between the 3DTI surface and thin-film FMM layer, respectively. Let us give detailed expressions for them in the following.

We start with the 3DTI term. In this paper, we consider \ce{Bi2Se3}-family as a concrete material of the 3DTI. Electronic states around the Fermi level and the \textGamma~point in these TIs are described by an effective tight-binding Hamiltonian derived in Ref. \cite{Liu-PRB82-045122}. The effective Hamiltonian forms the highest valence band and the lowest conduction band, both of which have two-fold degeneracy due to the presence of time-reversal symmetry and inversion symmetry. In bulk of the 3DTI, the effective Hamiltonian is explicitly written as
\begin{equation}
	\hat{\mathcal{H}}^{\mathrm{bulk}}_{\mathrm{TI}}=\sum_{\bsym{k} \in \mathrm{BZ}}c^{\dag}_{\bsym{k}}H^{\mathrm{bulk}}_{\mathrm{TI}}(\bsym{k})c_{\bsym{k}}, \label{eq:Hami_3DTI}
\end{equation}
where $c^{\dag}_{\bsym{k}}=
\left [ \begin{array}{cccc}
	\cre{c}_{\bsym{k}-\uparrow} & \cre{c}_{\bsym{k}-\downarrow} & \cre{c}_{\bsym{k}+\uparrow} & \cre{c}_{\bsym{k}+\downarrow}
\end{array} \right ]
$ is a 4-component spinor of creation operators of electrons in the 3DTI, and $H^{\mathrm{bulk}}_{\mathrm{TI}}(\bsym{k})$ is a $4 \times 4$ matrix given as, 
\begin{equation}
	H^{\mathrm{bulk}}_{\mathrm{TI}}(\bsym{k})=\epsilon_{\mathrm{TI}}(\bsym{k})1_{4}+\sum_{i=1}^{3}R_{i}(\bsym{k})\alpha_{i}+m(\bsym{k})\beta. \label{eq:HamiMat_3DTI}
\end{equation}
Here, $\bsym{k}=\left ( k_{x},k_{y},k_{z} \right )$ is the Bloch wave vector, $\uparrow$ ($\downarrow$) represents $z$ component of the total angular momentum $j_{z}=+1/2$ ($-1/2$), and $\pm$ means sign of the parity eigenvalue. In Eq. \eqref{eq:HamiMat_3DTI}, $1_{4}$ is the $4 \times 4$ identity matrix, and $\alpha_{i}$ and $\beta$ are Dirac matrices which can be explicitly written as
\begin{equation}
	\alpha_{i}=
	\left [ \begin{array}{cc}
		0 & \sigma_{i} \\
		\sigma_{i} & 0
    \end{array}\right ], \quad\beta=
    \left [ \begin{array}{cc}
		1_{2} & 0 \\
		0 & -1_{2}
    \end{array} \right ],
\end{equation}
where $\sigma_{i}$ are the Pauli matrices. We consider a 3DTI on a hexagonal lattice with nearest neighbor hoppings and thus the coefficients of these matrices in Eq. \eqref{eq:HamiMat_3DTI} can be written as
\begin{align}
	\epsilon_{\mathrm{TI}}(\bsym{k}) &= \epsilon_{0,\mathrm{TI}}+\frac{2\epsilon_{2}}{c^{2}}\left \{ 1-\cos\left [ k_{z}c \right ] \right \} \nonumber \\
	&\quad +\frac{4 \epsilon_{1}}{3a^{2}}\left \{ 3-2\cos\left [ \frac{\sqrt{3}}{2}k_{x}a \right ]\cos\left [ \frac{1}{2}k_{y}a \right ]-\cos\left [ k_{y}a \right ] \right \}, \\
	R_{1}(\bsym{k}) &= \frac{2A_{1}}{\sqrt{3}a}\sin\left [ \frac{\sqrt{3}}{2}k_{x}a \right ]\cos\left [ \frac{1}{2}k_{y}a \right ], \\
	R_{2}(\bsym{k}) &= \frac{2A_{1}}{3a}\left \{ \cos\left [ \frac{\sqrt{3}}{2}k_{x}a \right ]\sin\left [ \frac{1}{2}k_{y}a \right ]+\sin\left [ k_{y}a \right ] \right \}, \\
	R_{3}(\bsym{k}) &= A_{2}\sin\left [ k_{z}c \right ], \\
	m(\bsym{k}) &= m_{0}+\frac{2m_{2}}{c^{2}}\left \{ 1-\cos\left [ k_{z}c \right ] \right \} \nonumber \\
	&\quad +\frac{4m_{1}}{3a^{2}}\left \{ 3-2\cos\left [ \frac{\sqrt{3}}{2}k_{x}a \right ]\cos\left [ \frac{1}{2}k_{y}a \right ]-\cos\left [ k_{y}a \right ] \right \}
\end{align}
Here, $a$ and $c$ are lattice constants in $xy$ plane and along $z$ axis, respectively, and $m_{0}$ corresponds to the strength of the spin-orbit interaction. $a$, $c$, $\epsilon_{0,\mathrm{TI}}$, $\epsilon_{1}$, $\epsilon_{2}$, $A_{1}$, $A_{2}$, $m_{0}$, $m_{1}$, $m_{2}$ are material-dependent parameters. We determine these values by fitting the energy spectrum of this effective Hamiltonian to that of \ce{Bi2Se3} obtained by \textit{ab initio} calculation : $a=4.134~\left [ \si{\angstrom} \right ]$, $c=28.63~\left [ \si{\angstrom} \right ]$, $\epsilon_{1}=20.17~\left [ \si{eV}\si{\angstrom}^{2} \right ]$, $\epsilon_{2}=19.67~\left [ \si{eV}\si{\angstrom}^{2} \right ]$, $A_{1}=4.175~\left [ \si{eV}\si{\angstrom} \right ]$, $A_{2}=9.162~\left [ \si{eV}\si{\angstrom} \right ]$, $m_{0}=-0.28~\left [ \si{eV} \right ]$, $m_{1}=58.28~\left [ \si{eV}\si{\angstrom}^{2} \right ]$, $m_{2}=117.1~\left [ \si{eV}\si{\angstrom}^{2} \right ]$ \cite{Kurebayashi-JPSJ83-063709,Sasaki-PRL107-217001}. We took the value of $\epsilon_{0,\mathrm{TI}}$ as $\epsilon_{0,\mathrm{TI}}=-0.3111~\left [ \si{eV} \right ]$ so that the Dirac point on the 3DTI surface lies at the criterion of energy $E=0$.

From Eq. \eqref{eq:Hami_3DTI}, we implement a slab of the 3DTI with $N_{z}^{\mathrm{TI}}$ layers along the $z$ direction to construct the 3DTI/FMM bilayer structure. To obtain this open boundary geometry, we rewrite the $k_{z}$ dependence of the Eqs. \eqref{eq:Hami_3DTI} and \eqref{eq:HamiMat_3DTI} by truncating the hopping outside of $n_{z}=1,2, \ldots ,N^{\mathrm{TI}}_{z}$ layers in the tight-binding model. As a result of this procedure, the explicit form of $\hat{\mathcal{H}}_{\mathrm{TI}}$ in Eq. \eqref{eq:Hami_tot} is given as
\begin{equation}
	\hat{\mathcal{H}}_{\mathrm{TI}}=\sum_{k_{x},k_{y} \in \mathrm{SBZ}}c^{\dag}_{k_{x}k_{y}}H_{\mathrm{TI}}\left ( k_{x},k_{y} \right )c_{k_{x}k_{y}}, \label{eq:Hami_3DTI_Sur}
\end{equation}
where $c^{\dag}_{k_{x}k_{y}}=
\left [ \begin{array}{cccc}
	c^{\dag}_{k_{x}k_{y}1} & c^{\dag}_{k_{x}k_{y}2} & \ldots & c^{\dag}_{k_{x}k_{y}N^{\mathrm{TI}}_{z}}
\end{array} \right ]
$ and $ c^{\dag}_{k_{x}k_{y}n_{z}}=
\left [ \begin{array}{cccc}
	\cre{c}_{k_{x}k_{y}n_{z}-\uparrow} & \cre{c}_{k_{x}k_{y}n_{z}-\downarrow} & \cre{c}_{k_{x}k_{y}n_{z}+\uparrow} & \cre{c}_{k_{x}k_{y}n_{z}+\downarrow}
\end{array} \right ]
$ are matrices of the creation operators, SBZ means surface BZ, and
\begin{align}
	H_{\mathrm{TI}}\left ( k_{x},k_{y} \right ) &= 1_{N^{\mathrm{TI}}_{z}} \otimes \left \{ \tilde{\epsilon}_{\mathrm{TI}}\left ( k_{x},k_{y} \right )1_{4}+\sum_{i=1}^{2}R_{i}(\bsym{k})\alpha_{i} \right . \nonumber \\
	& \left . \quad\quad\quad\quad+\tilde{m}\left ( k_{x},k_{y} \right )\beta \right \}+H_{z}. \label{eq:HamiMat_3DTI_Sur-1}
\end{align}
Here, $\tilde{\epsilon}_{\mathrm{TI}}\left ( k_{x},k_{y} \right )=\epsilon_{\mathrm{TI}}(\bsym{k})+2B_{2}\cos\left [ k_{z}c \right ]/c^{2}$, $\tilde{m}\left ( k_{x},k_{y} \right )=m(\bsym{k})+2m_{2}\cos\left [ k_{z}c \right ]/c^{2}$, and
\begin{align}
	H_{z} &= 
	\left [ \begin{array}{cccc}
		0 & 1 &  &  \\
		1 & \ddots & \ddots &  \\
		& \ddots & \ddots & 1 \\
		&  & 1 & 0 \\
	\end{array} \right ] \otimes \left ( -\frac{B_{2}}{c^{2}}1_{4}-\frac{m_{2}}{c^{2}}\beta \right ) \nonumber \\
	&\quad +
	\left [ \begin{array}{cccc}
		0 & -\mi &  &  \\
		\mi & \ddots & \ddots &  \\
		& \ddots & \ddots & -\mi \\
		&  & \mi & 0 \\
	\end{array} \right ] \otimes \left ( \frac{A_{2}}{2c}\alpha_{3} \right ). \label{eq:HamiMat_3DTI_Sur-2}
\end{align}
In Eqs. \eqref{eq:HamiMat_3DTI_Sur-1} and \eqref{eq:HamiMat_3DTI_Sur-2}, the left-side matrices of the direct products have $N^{\mathrm{TI}}_{z} \times N^{\mathrm{TI}}_{z}$ size representing the degrees of freedom of finite layers in $z$ direction.

Next, we introduce the thin-film FMM term. As a simple model for this material, we employ a two-dimensional single-orbital tight-binding Hamiltonian \cite{Zhang-PRB94-014435,Hsu-PRB96-235433}. For simplicity, we implement the FMM in a two-dimensional hexagonal lattice whose lattice constant in $xy$ plane is the same as that of the 3DTI. The tight-binding Hamiltonian is explicitly written as
\begin{equation}
	\hat{\mathcal{H}}_{\mathrm{FMM}}=\sum_{k_{x},k_{y} \in \mathrm{SBZ}}d^{\dag}_{k_{x}k_{y}n_{z}}H_{\mathrm{FMM}}\left ( k_{x},k_{y} \right )d_{k_{x}k_{y}n_{z}},
\end{equation}
where $d^{\dag}_{k_{x}k_{y}n_{z}}=
\left [ \begin{array}{cc}
    \cre{d}_{k_{x}k_{y}n_{z}\uparrow} & \cre{d}_{k_{x}k_{y}n_{z}\downarrow}
\end{array}\right ]
$ is a 2-component spinor of electron in thin-film FMM of $n_{z}$-th layer, and
\begin{align}
	H_{\mathrm{FMM}}\left ( k_{x},k_{y} \right ) &= \epsilon_{\mathrm{FMM}}\left ( k_{x},k_{y} \right )1_{2}-J\bsym{M} \cdot \bsym{\sigma}+\lambda_{\mathrm{R}}\left \{ \bsym{r}(\bsym{k}) \times \bsym{\sigma} \right \} \cdot \bsym{e}_{z}, \label{eq:HamiMat_FMM} \\
	\epsilon_{\mathrm{FMM}}\left ( k_{x},k_{y} \right ) &= \epsilon_{0,\mathrm{FMM}}-t\left \{ \cos\left [ \frac{\left ( \sqrt{3}k_{x}-k_{y} \right )a}{2} \right ] \right . \nonumber \\
	&\quad \left . +\cos\left [ \frac{\left ( \sqrt{3}k_{x}+k_{y} \right )a}{2} \right ]+\cos\left [ k_{y}a \right ] \right \}. \label{eq:HamiMat_FMM_TB}
\end{align}
In Eq. \eqref{eq:HamiMat_FMM}, the first term is the hopping term, the second term is the exchange interaction between magnetization $\bsym{M}$ and electron's spin, and the last term is the Rashba spin-orbit interaction which reflects two-dimensional character of FMM. Here we introduce vectors of $\bsym{r}(\bsym{k})=\left ( R_{1}(\bsym{k})/A_{1},R_{2}(\bsym{k})/A_{1},R_{3}(\bsym{k})/A_{2} \right )$ and $\bsym{\sigma}=\left ( \sigma_{1},\sigma_{2},\sigma_{3} \right )$. $\bsym{e}_{z}$ is the unit vector in $z$ direction. $\epsilon_{0,\mathrm{FMM}}$, $t$, $\lambda_{\mathrm{R}}$ and $J$ are the onsite energy, the hopping integral, the Rashba constant and the exchange interaction constant, respectively. By adjusting these parameters, we change the position of FMM band to investigate the requirement for wide QAH gap in Sec. \ref{sec:Condition}.

Then, we define the interface-hopping term. We locate the two-dimensional FMM on both sides of the 3DTI as shown in Fig. \ref{fig:System}, i.e., we set the thin-film FMM layers at $n_{z}=0,N_{z}^{\mathrm{TI}}+1$ layers. When the FMM is attached to the surface of 3DTI, interface coupling emerges between the $n_{z}=0$ ($N_{z}^{\mathrm{TI}}+1$) layer and the $n_{z}=1$ ($N_{z}^{\mathrm{TI}}$) layer. Thus we explicitly model the interface-hopping term as
\begin{align}
	\hat{\mathcal{H}}_{\mathrm{IH}} &= t_{\mathrm{IH}}\sum_{k_{x},k_{y} \in \mathrm{SBZ}}\left ( d^{\dag}_{k_{x}k_{y}0}
	\left [ \begin{array}{cc}
		1_{2} & 1_{2}
	\end{array} \right ]
	c_{k_{x}k_{y}1} \right . \nonumber \\
	&\quad \left . +d^{\dag}_{k_{x}k_{y}N_{z}^{\mathrm{TI}}+1}
	\left [ \begin{array}{cc}
		1_{2} & 1_{2}
	\end{array} \right ]
	c_{k_{x}k_{y}N_{z}^{\mathrm{TI}}} \right )+\mathrm{H.~c.}, \label{eq:Hami_IH}
\end{align}
where $t_{\mathrm{IH}}$ is the interface-hopping integral. Here, for simplicity, we assume that the interface coupling is independent of the orbital and spin, and non-zero only between the states with the same angular momentum.

Finally, we can rewrite the Eq. \eqref{eq:Hami_tot} in matrix form as
\begin{align}
	\hat{\mathcal{H}} &= \sum_{k_{x},k_{y} \in \mathrm{SBZ}}
	\left [ \begin{array}{ccc}
		d^{\dag}_{k_{x}k_{y}0} & c^{\dag}_{k_{x}k_{y}} & d^{\dag}_{k_{x}k_{y}N_{z}^{\mathrm{TI}}+1}
	\end{array} \right ] \nonumber \\
	&\quad \times\left [ \begin{array}{ccc}
		H_{\mathrm{FMM}}\left ( k_{x},k_{y} \right ) & H_{\mathrm{IH}} &  \\
		{H_{\mathrm{IH}}}^{\dag} & H_{\mathrm{TI}}\left ( k_{x},k_{y} \right ) & {H_{\mathrm{IH}}}^{\dag} \\
		& H_{\mathrm{IH}} & H_{\mathrm{FMM}}\left ( k_{x},k_{y} \right )
	\end{array} \right ]
	\left [ \begin{array}{c}
		d_{k_{x}k_{y}0} \\
		c_{k_{x}k_{y}} \\
		d_{k_{x}k_{y}N_{z}^{\mathrm{TI}}+1}
	\end{array} \right ] \nonumber \\
	&= \sum_{k_{x},k_{y} \in \mathrm{SBZ}}
	\left [ \begin{array}{ccc}
		d^{\dag}_{k_{x}k_{y}0} & c^{\dag}_{k_{x}k_{y}} & d^{\dag}_{k_{x}k_{y}N_{z}^{\mathrm{TI}}+1}
	\end{array} \right ]H\left ( k_{x},k_{y} \right )\left [ \begin{array}{c}
		d_{k_{x}k_{y}0} \\
		c_{k_{x}k_{y}} \\
		d_{k_{x}k_{y}N_{z}^{\mathrm{TI}}+1}
	\end{array} \right ], \label{eq:HamiMat_tot}
\end{align}
where $H_{\mathrm{IH}}=t_{\mathrm{IH}}
\left [ \begin{array}{cc}
	1_{2} & 1_{2}
\end{array} \right ]$.

\section{Results} \label{sec:Result}
\subsection{Topological proximity effect in 3DTI/NM bilayer} \label{sec:TPE}
\begin{figure*}[tbhp]
	\centering
	\begin{subfigure}[t]{0.22\textwidth}
		\centering
		\caption{$t_{\mathrm{IH}}=0~\left [ \si{eV} \right ]$}
		\includegraphics{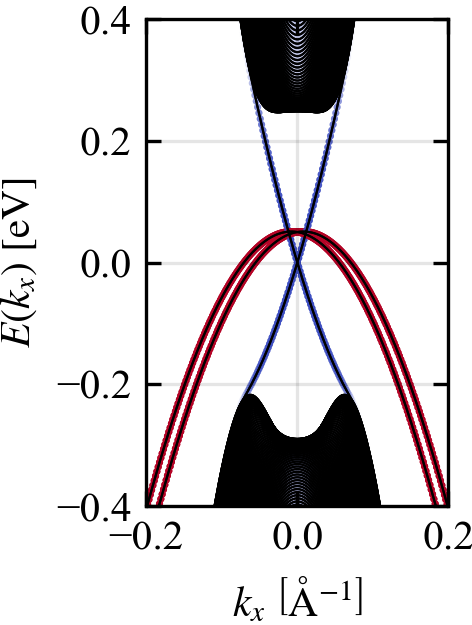}
		\label{fig:TPE_Band_t=0}
	\end{subfigure}
	\begin{subfigure}[t]{0.22\textwidth}
		\centering
		\caption{$t_{\mathrm{IH}}=0.1~\left [ \si{eV} \right ]$}
		\includegraphics{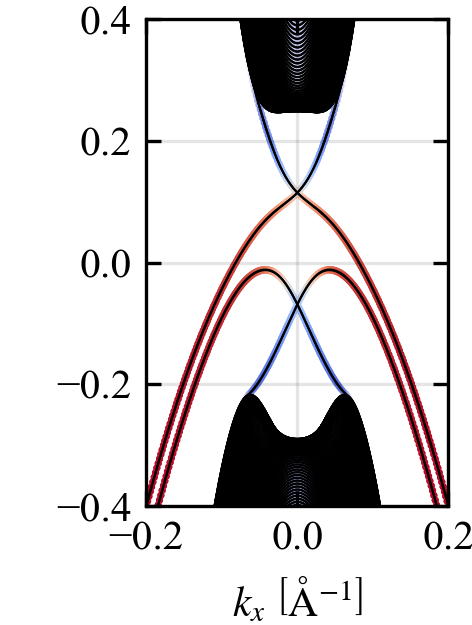}
		\label{fig:TPE_Band_t=0.1}
	\end{subfigure}
	\begin{subfigure}[t]{0.22\textwidth}
		\centering
		\caption{$t_{\mathrm{IH}}=0.3~\left [ \si{eV} \right ]$}
		\includegraphics{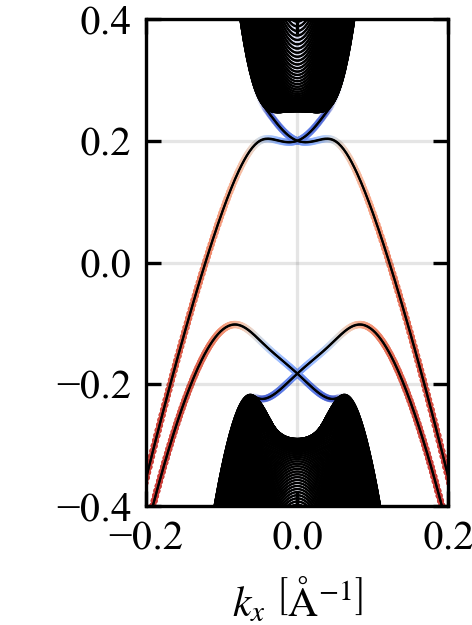}
		\label{fig:TPE_Band_t=0.3}
	\end{subfigure}
	\begin{subfigure}[t]{0.3\textwidth}
		\centering
		\caption{}
		\includegraphics[keepaspectratio, scale=0.5]{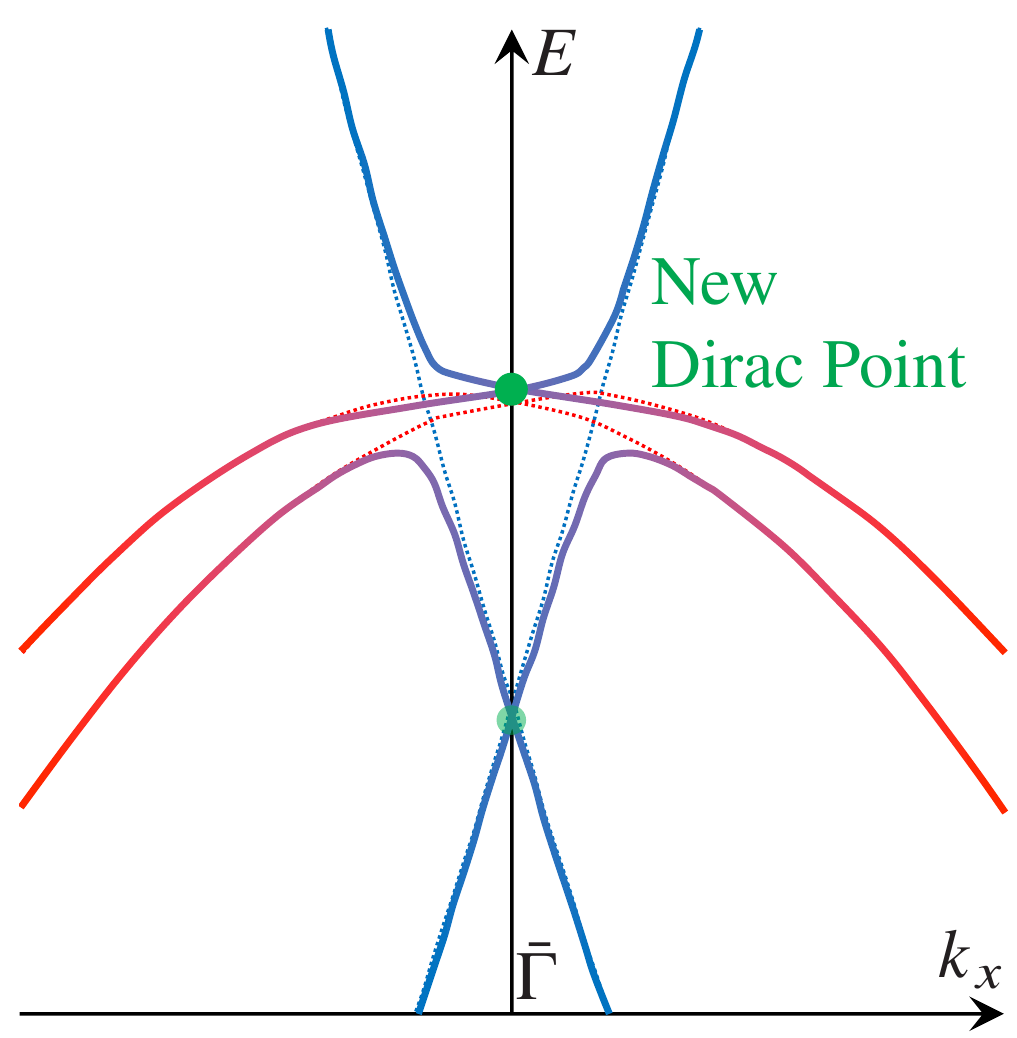}
		\label{fig:Hybridization}
	\end{subfigure} \\
	\begin{subfigure}[t]{0.22\textwidth}
		\centering
		\caption{$t_{\mathrm{IH}}=0~\left [ \si{eV} \right ]$}
		\includegraphics{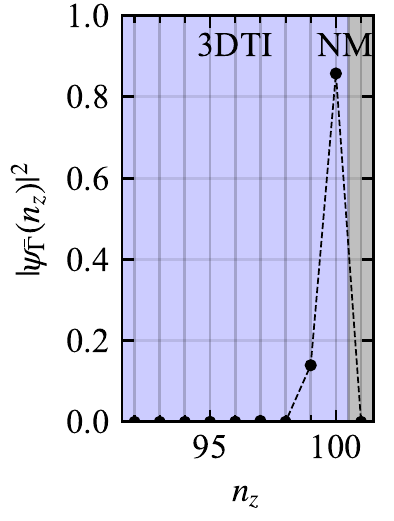}
		\label{fig:TPE_EgVec_t=0}
	\end{subfigure}
	\begin{subfigure}[t]{0.22\textwidth}
		\centering
		\caption{$t_{\mathrm{IH}}=0.1~\left [ \si{eV} \right ]$}
		\includegraphics{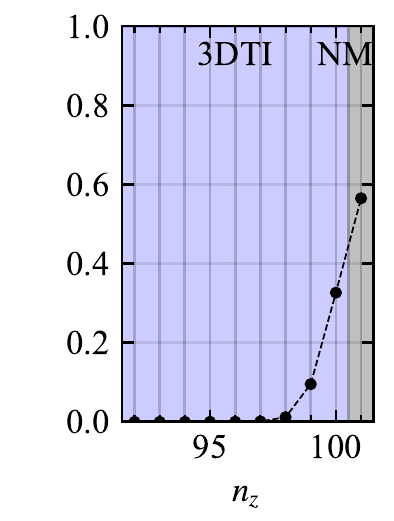}
		\label{fig:TPE_EgVec_t=0.1}
	\end{subfigure}
	\begin{subfigure}[t]{0.22\textwidth}
		\centering
		\caption{$t_{\mathrm{IH}}=0.3~\left [ \si{eV} \right ]$}
		\includegraphics{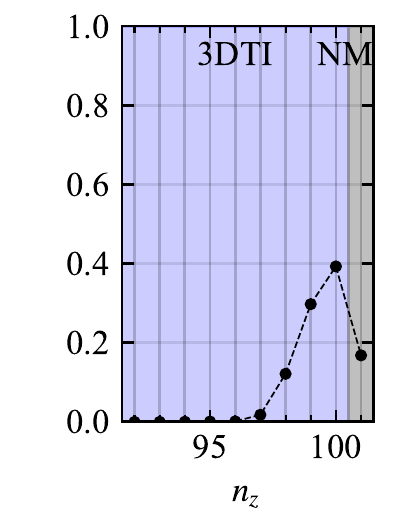}
		\label{fig:TPE_EgVec_t=0.3}
	\end{subfigure}
	\begin{subfigure}[t]{0.3\textwidth}
		\centering
		\caption{}
		\includegraphics{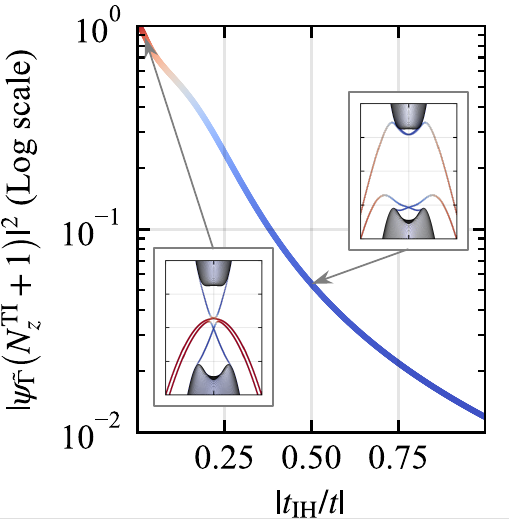}
		\label{fig:TPE_EgVec_Hop}
	\end{subfigure}
	\caption{\subref{fig:TPE_Band_t=0}-\subref{fig:TPE_Band_t=0.3} Band structures of $H\left ( k_{x},k_{y} \right )$ on the $k_{y}=0$ line, for different interface-hopping values. Red (blue)-colored plots on the energy band represents that the weight by Eq. \eqref{eq:WeightDistribution} is large at FMM (3DTI surface) layers. The vertical axes of these figures have their scales in common. \subref{fig:Hybridization} Schematic picture describing the shift of gapless Dirac point due to the hybridization breaking the band intersection. The color of lines corresponds to the spin direction shown at the bottom of figure. \subref{fig:TPE_EgVec_t=0}-\subref{fig:TPE_EgVec_t=0.3} Weight distributions of the topologically protected Dirac points, calculated with Eq. \eqref{eq:WeightDistribution} and for different interface-hopping values. The blue (gray) area represents the layers of the 3DTI (NM). The vertical axes of these figures have their scales in common. \subref{fig:TPE_EgVec_Hop} Variation of the weight of the Dirac point in NM layer as a function of interface-hopping strength within the range of $0<\left | t_{\mathrm{IH}}/t \right |<1$. Band structures calculated at certain values of $\left | t_{\mathrm{IH}}/t \right |$ are shown in the insets, whose axes have the same quantities and scales as Figs. \subref{fig:TPE_Band_t=0}-\subref{fig:TPE_Band_t=0.3}. We took the adjustable parameters in Eqs. \eqref{eq:HamiMat_FMM} and \eqref{eq:HamiMat_FMM_TB} as $\epsilon_{0}=-2.95~\left [ \si{eV} \right ]$, $t=-1.0~\left [ \si{eV} \right ]$, $J=0~\left [ \si{eV} \right ]$, and $\lambda_{\mathrm{R}}=0.2~\left [ \si{eV} \right ]$.}
	\label{fig:TPE}
\end{figure*}
Before discussing the 3DTI/thin-film FMM geometry, here we consider 3DTI/thin-film NM bilayer structure, that is, the case of $J=0$ in Eq. \eqref{eq:HamiMat_FMM}. Under this condition, we numerically diagonalize the Hamiltonian $H\left ( k_{x},k_{y} \right )$ of Eq. \eqref{eq:HamiMat_tot}, and obtain eigenvalues $E_{n}\left ( k_{x},k_{y} \right )$ and $4N_{z}^{\mathrm{TI}}+4$-component eigenvectors $\bsym{u}_{n}\left ( k_{x},k_{y} \right )$, where $n=1,2, \ldots ,4N_{z}^{\mathrm{TI}}+4$ is index of energy bands. From the eigenvectors, we can calculate the layer-resolved weight distribution of electron density corresponding to each eigenvalue as
\begin{equation}
	\left | \psi_{nk_{x}k_{y}}\left ( n_{z} \right ) \right |^{2}=
	\begin{cases}
		\displaystyle \sum_{i=1}^{2}\left | u_{n}^{(i)}\left ( k_{x},k_{y} \right ) \right |^{2} & \left ( n_{z}=0 \right ) \\
		\displaystyle \sum_{i=4n_{z}-1}^{4n_{z}+2}\left | u_{n}^{(i)}\left ( k_{x},k_{y} \right ) \right |^{2} & \left ( 1 \leq n_{z} \leq N_{z}^{\mathrm{TI}} \right ) \\
		\displaystyle \sum_{i=4N^{\mathrm{TI}}_{z}+3}^{4N^{\mathrm{TI}}_{z}+4}\left | u_{n}^{(i)}\left ( k_{x},k_{y} \right ) \right |^{2} & \left ( n_{z}=N_{z}^{\mathrm{TI}}+1 \right )
    \end{cases}, \label{eq:WeightDistribution}
\end{equation}
where $u_{n}^{(i)}\left ( k_{x},k_{y} \right )$ represents $i$-th component of the eigenvector. Figs. \bothref{fig:TPE}{fig:TPE_Band_t=0}-\bothref{fig:TPE}{fig:TPE_EgVec_t=0.3} show the band structures and the layer-resolved weight distributions of topologically non-trivial gapless state at the $\bar{\Gamma}$ point for different values of $t_{\mathrm{IH}}$. The gapless state connects the valence band of bulk 3DTI or NM and the conduction band. In Figs. \bothref{fig:TPE}{fig:TPE_Band_t=0}-\bothref{fig:TPE}{fig:TPE_Band_t=0.3}, the color gradient represents the projected weight on the surface layers of 3DTI (blue) or the NM layers (red) for each eigenstate. Here, we took the adjustable parameters in Eqs. \eqref{eq:HamiMat_FMM} and \eqref{eq:HamiMat_FMM_TB} as $\epsilon_{0}=-2.95~\left [ \si{eV} \right ]$, $t=-1.0~\left [ \si{eV} \right ]$ and $\lambda_{\mathrm{R}}=0.2~\left [ \si{eV} \right ]$ so that the original NM band and the linear dispersion of 3DTI intersect. 

At first, we study the case of $t_{\mathrm{IH}}=0~\left [ \si{eV} \right ]$ where the 3DTI and the thin-film NM are spatially separated. In this setting, the Hamiltonian of Eq. \eqref{eq:HamiMat_tot} can be individually diagonalized at the 3DTI part and the NM part, and we obtain independent topologically trivial NM band (red lines) and non-trivial linear dispersion (blue lines) as shown in Fig. \bothref{fig:TPE}{fig:TPE_Band_t=0}. Here, we determine the band connecting the conduction band of 3DTI and the valence bands of 3DTI or NM as topologically non-trivial. At the $\bar{\Gamma}$ point, the Dirac dispersion on 3DTI surface forms topologically protected Dirac point. The band structure of 3DTI around $k_{x}=0$ agrees well with the one obtained by \textit{ab initio} calculation in Ref. \cite{Zhang-NatPhys5-438}. The weight distribution of the eigenstate at the Dirac point has a sharp maximum at the top layer of 3DTI surface and has no amplitude in NM layer as Fig. \bothref{fig:TPE}{fig:TPE_EgVec_t=0} shows.

Next, we introduce non-zero value for the interface-hopping strength of $t_{\mathrm{IH}}=0.1~\left [ \si{eV} \right ]$. Resulting band structure is shown in Fig. \bothref{fig:TPE}{fig:TPE_Band_t=0.1}. In the figure, one can see that the topologically non-trivial Dirac dispersion appears above the topologically trivial band, which is in sharp contrast to the case of $t_{\mathrm{IH}}=0$. A mechanism of the formation of these bands can be understood as follows. Let us start with the situation of Fig. \bothref{fig:TPE}{fig:TPE_Band_t=0}. Once the 3DTI and the thin-film NM are attached together, non-zero $t_{\mathrm{IH}}$ appears in non-diagonal part of Eq. \eqref{eq:HamiMat_tot} and hybridizes the Dirac dispersion on 3DTI surface and the NM band which intersect each other at some points. As schematically shown in Fig. \bothref{fig:TPE}{fig:Hybridization}, such a hybridization between the intersecting bands breaks the degenerated points. As a result, the original topologically non-trivial (trivial) band is pushed up (down) to reconstruct a distorted Dirac dispersion. Such deformation of energy bands has been experimentally established in Ref. \cite{Shoman-NatCommun6-6547}. To observe the eigenstates of the reconstructed topologically-non-trivial Dirac dispersion in detail, we show layer-resolved weight distribution of the topologically non-trivial Dirac point in Fig. \bothref{fig:TPE}{fig:TPE_EgVec_t=0.1}. The weight has a maximum at the thin-film NM layer, which indicates that the Dirac electrons migrate from the 3DTI surface into the NM layer due to the attachment of the thin-film NM.

Then, we discuss the case where larger interface coupling of $t_{\mathrm{IH}}=0.3~\left [ \si{eV} \right ]$ exists, as shown in Fig. \bothref{fig:TPE}{fig:TPE_Band_t=0.3}. Comparing with Fig. \bothref{fig:TPE}{fig:TPE_Band_t=0.1}, one can easily see that stronger interface hopping leads to a larger split between the energy bands within the bulk gap of 3DTI. Because of this large split, the Dirac point of topologically non-trivial band shifts close to the conduction band of bulk 3DTI. As a result, the weight distribution at the point moves back to the 3DTI as Fig. \bothref{fig:TPE}{fig:TPE_EgVec_t=0.3} shows. To see a continuous variation of the weight distribution with respect to the interface-coupling strength, we plot the weight in NM layer at the Dirac point as a function of $t_{\mathrm{IH}}$ as shown in Fig. \bothref{fig:TPE}{fig:TPE_EgVec_Hop}. As this figure shows, the topologically stable Dirac states strongly localize at NM layer in weak-attached limit, but it rapidly penetrates back to the 3DTI as the value of $t_{\mathrm{IH}}$ becomes large. This is because strong interface coupling widely breaks the intersect between the linear dispersion of 3DTI and the NM band and energetically pushes up the reconstructed Dirac cone. The pushed-up Dirac cone incorporates with the valence band of bulk 3DTI and eigenstates of the Dirac cone possess wide weight in the 3DTI.

At the end of this section, we discuss the ideal condition of the topological proximity effect and decide the value of $t_{\mathrm{IH}}$ which we use in the later sections. To realize a wide band gap in the reconstructed topologically-non-trivial Dirac dispersion by breaking the time-reversal symmetry, the topologically non-trivial Dirac dispersion should be separated from the bands of bulk 3DTI and other surface dispersion. In this paper, we have set the NM band right above the original Dirac point of the linear dispersion on 3DTI surface as Fig. \bothref{fig:TPE}{fig:TPE_Band_t=0} so that both energy bands intersect close to the Dirac point. In this situation, from Fig. \bothref{fig:TPE}{fig:TPE_Band_t=0.1} and \bothref{fig:TPE}{fig:TPE_EgVec_Hop}, we can see that the reconstructed Dirac dispersion is isolated from both the other reconstructed topologically-trivial Dirac band and the conduction band of bulk 3DTI at around $\left | t_{\mathrm{IH}}/t \right |=0.1$. The topologically non-trivial Dirac cone is located close to the topologically trivial one (the conduction band of bulk 3DTI) when $\left | t_{\mathrm{IH}}/t \right |<0.1$ ($\left | t_{\mathrm{IH}}/t \right |>0.1$) as shown in the insets of Fig. \bothref{fig:TPE}{fig:TPE_EgVec_Hop}. Therefore, we fix the value of interface hopping strength as $\left | t_{\mathrm{IH}}/t \right |=0.1$ in the later sections as the ideal condition for the emergence of wide band gap. In this setting, the eigenstate of the topologically non-trivial Dirac point migrates into the NM layer as we have shown in Fig. \bothref{fig:TPE}{fig:TPE_EgVec_t=0.1}.

\subsection{Quantum anomalous Hall effect in 3DTI/FMM bilayer} \label{sec:QAHE}
\begin{figure}[tbhp]
	\centering
	\includegraphics{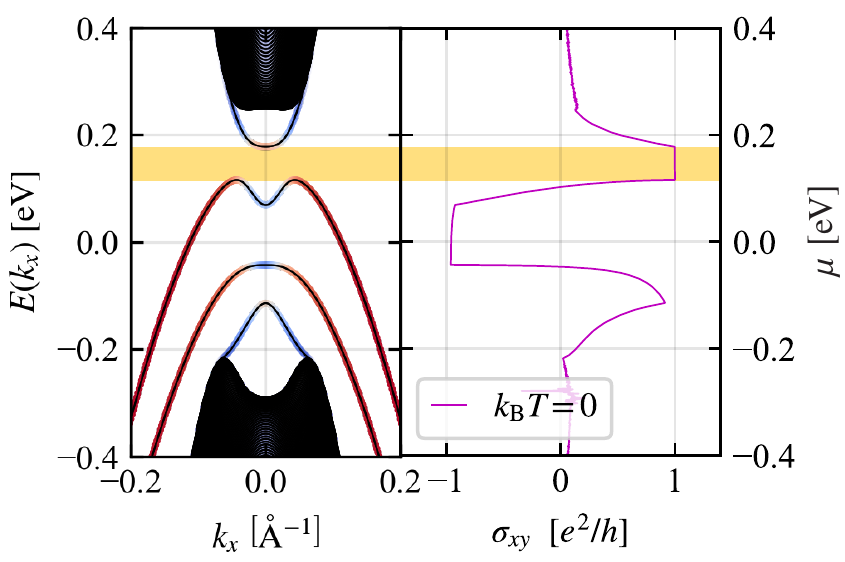}
	\caption{Band structure of the 3DTI/thin-film FMM bilayer structure at left side, and corresponding anomalous Hall conductivity as a function of the chemical potential at right side. We took the adjustable parameters in Eqs. \eqref{eq:HamiMat_FMM}, \eqref{eq:HamiMat_FMM_TB} and \eqref{eq:Hami_IH} as $\epsilon_{0}=-2.95~\left [ \si{eV} \right ]$, $t=-1.0~\left [ \si{eV} \right ]$, $JM_{z}=0.1~\left [ \si{eV} \right ]$ with $\bsym{M}=M_{z}\bsym{e}_{z}$, $\lambda_{\mathrm{R}}=0.2~\left [ \si{eV} \right ]$ and $t_{\mathrm{IH}}=0.1~\left [ \si{eV} \right ]$.}
	\label{fig:BandHC_0and0}
\end{figure}
Here we turn on the exchange interaction between the uniform magnetization and electron spin in thin-film metal layers to study a 3DTI/thin-film FMM bilayer. In addition to the band structure, we compute the anomalous Hall conductivity $\sigma_{xy}$ from intrinsic mechanism using the Kubo formula,
\begin{equation}
	\sigma_{xy}=\frac{e^{2}}{h}\sum_{n}\int_{\mathrm{SBZ}}\frac{\md k_{x}\md k_{y}}{2\mpi}b^{(z)}_{n}\left ( k_{x},k_{y} \right )f(E_{n}-\mu), \label{eq:HC}
\end{equation}
where $n$ is the band index, $\mu$ is the chemical potential of whole system and $f(E_{n}-\mu)$ is the Fermi-Dirac distribution function. $b^{(z)}_{n}\left ( k_{x},k_{y} \right )=\partial_{k_{x}}a^{(y)}_{n}\left ( k_{x},k_{y} \right )-\partial_{k_{y}}a^{(x)}_{n}\left ( k_{x},k_{y} \right )$ ($\partial_{k_{\eta}} \equiv \partial/\partial k_{\eta}$ for $\eta=x,y$) is $z$-component of the Berry curvature, where $a^{(\eta)}_{n}\left ( k_{x},k_{y} \right )=-\mi\Braket{u_{n}\left ( k_{x},k_{y} \right )|\partial_{k_{\eta}}u_{n}\left ( k_{x},k_{y} \right )}$ is the Berry connection and $\Ket{u_{n}\left ( k_{x},k_{y} \right )}$ is the Bloch eigenstate obtained by the diagonalization. 

The computed band structure is shown at the left side of Fig. \ref{fig:BandHC_0and0}. Here, we set the magnetization vector in thin-film FMM to be perpendicular to the surface, i.e., $\bsym{M}=M_{z}\bsym{e}_{z}$ (we discuss reasonability of this setting in Sec. \ref{sec:MagneticAnisotropy}), and took the value of exchange-interaction strength in FMM as $JM_{z}=0.1~\left [ \si{eV} \right ]$. Comparing the band structure of Fig. \ref{fig:BandHC_0and0} with Fig. \bothref{fig:TPE}{fig:TPE_Band_t=0.1}, one can see that non-zero exchange interaction in FMM opens surface gap to the Dirac points at $\bar{\Gamma}$ point. Remarkably, the width of the surface gap is as wide as about $60~\left [ \si{meV} \right ]$, which is larger than the magnetically-doped thin-film TIs, though the exchange-interaction strength is much smaller than the typical values of that in actual transition ferromagnetic metals. In Sec. \ref{sec:Condition}, we study a condition for the appearance of this wide surface gap.

At the right side of Fig. \ref{fig:BandHC_0and0}, we show the corresponding anomalous Hall conductivity at the temperature $T=0$ as a function of the chemical potential. Since the global surface gap exists over the whole SBZ, the 3DTI/FMM bilayer structure possesses a constant quantized anomalous Hall conductivity when the chemical potential lies in the range of the surface gap. This quantization can be qualitatively understood by considering how the exchange interaction works to the Dirac electrons migrating into the FMM layers. Let us focus on one side of the thin-film FMM. We adopt a simple $2 \times 2$ Dirac Hamiltonian,
\begin{equation}
	H_{\mathrm{2D}}\left ( k_{x},k_{y} \right )=\hbar v\left ( \bsym{\sigma} \times \bsym{k} \right ) \cdot \bsym{e}_{z},
\end{equation}
to describe the two-dimensional spin-momentum locked eigenstates around the new Dirac point shown in Fig. \bothref{fig:TPE}{fig:Hybridization}. Here, $v$ is a constant which has dimension of velocity. The exchange interaction for the Dirac electrons is described as $H_{\mathrm{ex}}=-JM_{z}\sigma_{z}$, which works as a mass term for the Dirac Hamiltonian. Thus the exchange interaction opens a mass gap to the Dirac dispersion as shown at the left side of Fig. \ref{fig:BandHC_0and0}. Assuming that the chemical potential is in the gap, one can easily calculate the Berry curvature and the anomalous Hall conductivity of \eqref{eq:HC} from $H_{\mathrm{2D}}\left ( k_{x},k_{y} \right )+H_{\mathrm{ex}}$ as
\begin{equation}
	\sigma_{xy}=\mathrm{sgn}\left [ JM_{z} \right ]\frac{e^{2}}{2h},
\end{equation}
at $T=0$. This Hall conductivity is a contribution from one side of the FMM layers, and the one from the other side is calculated similarly. Therefore, we obtain the quantized Hall conductivity shown in Fig. \ref{fig:BandHC_0and0} as contribution from the whole of 3DTI/FMM bilayer structure. 

\subsection{Magnetic anisotropy of the 3DTI/FMM system} \label{sec:MagneticAnisotropy}
\begin{figure}[tbhp]
	\centering
	\begin{subfigure}[t]{0.2\textwidth}
		\centering
		\caption{}
		\includegraphics[keepaspectratio, scale=0.5]{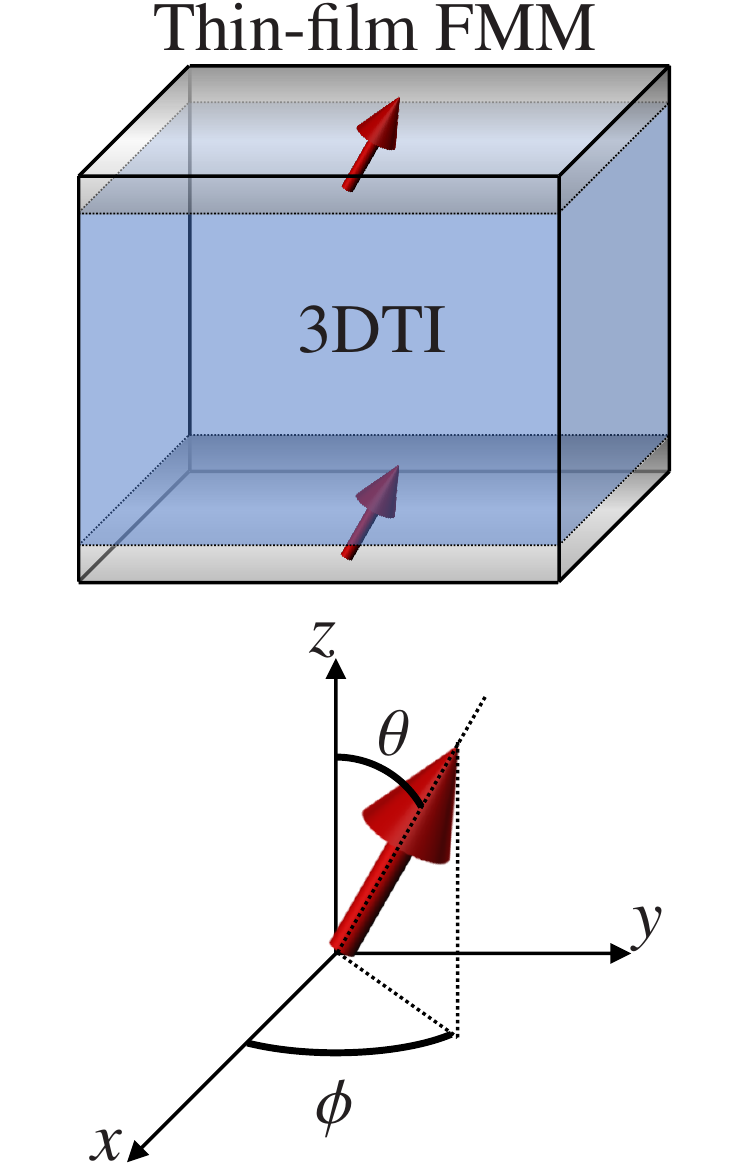}
		\label{fig:Angle}
	\end{subfigure}
	\begin{subfigure}[t]{0.25\textwidth}
		\centering
		\caption{}
		\includegraphics{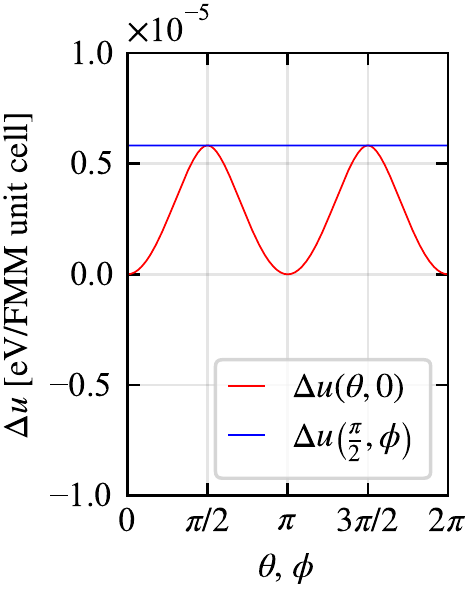}
		\label{fig:TotalEnergy}
	\end{subfigure} \\
	\begin{subfigure}[t]{0.23\textwidth}
		\centering
		\caption{$(\theta,\phi)=(0,0)$}
		\includegraphics{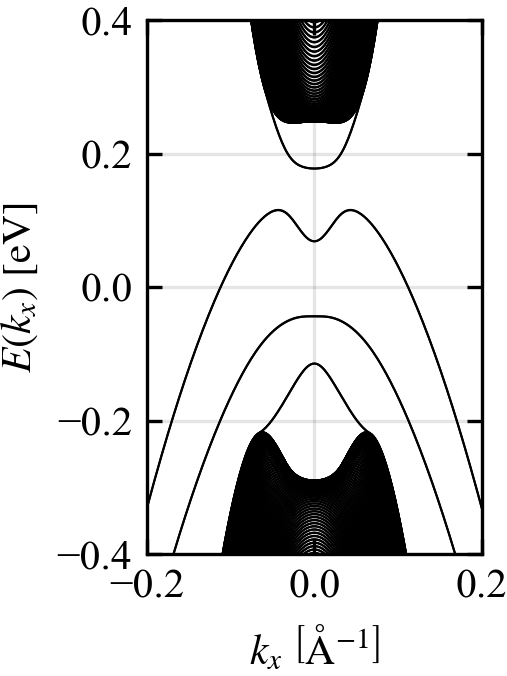}
		\label{fig:MA_theta=0}
	\end{subfigure}
	\begin{subfigure}[t]{0.23\textwidth}
		\centering
		\caption{$(\theta,\phi)=(\mpi/2,0)$}
		\includegraphics{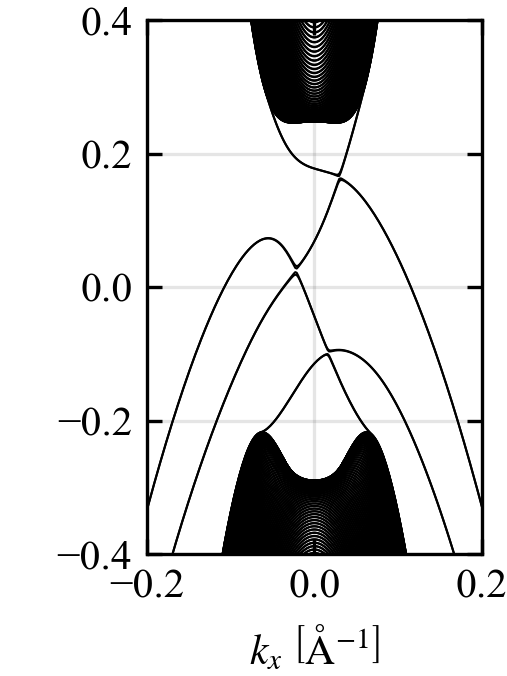}
		\label{fig:MA_theta=pi/2}
	\end{subfigure}
	\caption{\subref{fig:Angle} Definition of two angles representing the tilting of magnetization, $\theta$ and $\phi$. \subref{fig:TotalEnergy} Relative variation $\Delta u=u(\theta,\phi)-u(\theta=\phi=0)$ of Eq. \eqref{eq:TotalEnergy} as a function of $\theta$ and $\phi$. \subref{fig:MA_theta=0}, \subref{fig:MA_theta=pi/2} Band structure when the magnetization angle points \subref{fig:MA_theta=0} out-of-plane direction and \subref{fig:MA_theta=pi/2} in-plane direction against the surface. The vertical axes of these figures have their scales in common.}
	\label{fig:MagneticAnisotropy}
\end{figure}
Here we study magnetic anisotropy of the bilayer structure and discuss possibility of the appearance of QAH effect. To study the anisotropy, we compute the total energy of electrons in the system,
\begin{equation}
	u(T,\bsym{M})=\frac{1}{N_{\mathrm{FMM}}}\sum_{n}\int_{\mathrm{SBZ}}\md k_{x}\md k_{y}E_{n}\left ( k_{x},k_{y} \right )f(E_{n}-\mu), \label{eq:TotalEnergy}
\end{equation}
at $T=0$. Here, we divide the total energy by $N_{\mathrm{FMM}}$, the total number of unit cells in two FMM layers, so that we compute the intensive energy of magnetic anisotropy per one magnetized unit cell. The magnetization vector is expressed with polar coordinates as $\bsym{M}(\theta,\phi)=|\bsym{M}|\left ( \sin\theta\cos\phi,\sin\theta\sin\phi,\cos\theta \right )$ where $\theta$ and $\phi$ are defined as Fig. \bothref{fig:MagneticAnisotropy}{fig:Angle}. We fix $|\bsym{M}|$ with $J$ to be $J|\bsym{M}|=0.1~\left [ \si{eV} \right ]$, the same value as Sec. \ref{sec:QAHE}. The chemical potential $\mu$ in Eq. \eqref{eq:TotalEnergy} is determined so that the valence bands below the global surface gap in Fig. \ref{fig:BandHC_0and0} is completely filled at $T=0$. We derive relative variation from Eq. \eqref{eq:TotalEnergy} as a function of $\theta$ and $\phi$ independently, taking the value of $u(\theta=\phi=0)$ as the criterion for the total energy. Fig. \bothref{fig:MagneticAnisotropy}{fig:TotalEnergy} shows the calculated relative variation $\Delta u$. From the figure, one can easily see that the total energy as a function of $\theta$, $\Delta u(\theta,0)$, has minima at $\theta=0$ and $\mpi$, while a function of $\phi$, $\Delta u(\mpi/2,\phi)$, is constant for all $\phi$ values in the range of $0 \leq \phi \leq 2\mpi$. These behaviors clearly indicate that the easy axis of the magnetization in the thin-film FMM is perpendicular to the surface. In other words, once macroscopic magnetization appears in the FMM layers at a finite temperature, it points to the direction perpendicular to the surface, and the system can become QAH insulator with wide surface gap.

This character of magnetic anisotropy can be understood by considering topological properties of band structure as follows. As discussed in Sec. \ref{sec:QAHE}, the surface gap appearing as Fig. \bothref{fig:MagneticAnisotropy}{fig:MA_theta=0} corresponds to the topologically-stable quantized anomalous Hall conductivity of $\sigma_{xy}=e^{2}/2h$ for both the top and the bottom surfaces. Similarly, for $\theta=\mpi$, the surface gap opens with $\sigma_{xy}=-e^{2}/2h$ for each surface. The Hall conductivity does not depend on any parameters of the system, and thus can be interpreted to be a topological number as a unit of a physical constant $e^{2}/h$. These situations of $\theta=0$ and $\theta=\mpi$ accord each other by adiabatically rotate the magnetization vector. Therefore, the topological number should jump from one to the other along the adiabatic rotation, indicating that the global surface gap should close during the adiabatic change. To close the surface gap, the filled surface valence band adiabatically shifts upward to touch the empty surface conduction band as shown in Fig. \bothref{fig:MagneticAnisotropy}{fig:MA_theta=pi/2}, and the energy levels of electrons occupying the surface valence band increase. As a result, the total energy of electrons increases by rotating the magnetization in FMM layers to the in-plane direction, and therefore the magnetization vector prefers to point to the direction perpendicular to the surface.

\subsection{Condition to the wide quantum anomalous Hall gap} \label{sec:Condition}
\begin{figure}[tbhp]
	\centering
	\begin{subfigure}[t]{0.45\textwidth}
		\centering
		\caption{}
		\includegraphics[keepaspectratio]{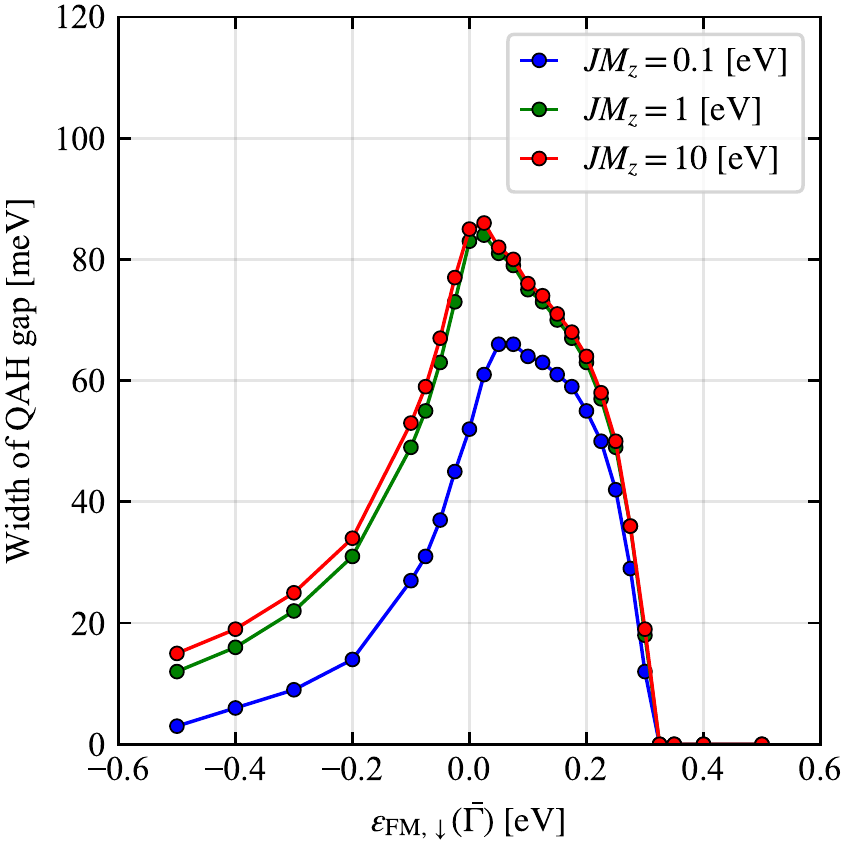}
		\label{fig:SurfaceGap}
	\end{subfigure} \\
	\begin{subfigure}[t]{0.23\textwidth}
		\centering
		\caption{}
		\includegraphics{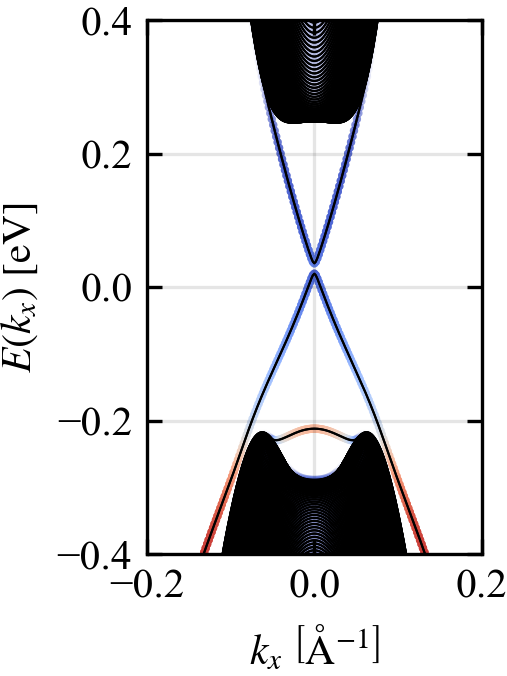}
		\label{fig:SG_Ins_Band_t=0.1}
	\end{subfigure}
	\begin{subfigure}[t]{0.23\textwidth}
		\centering
		\caption{}
		\includegraphics{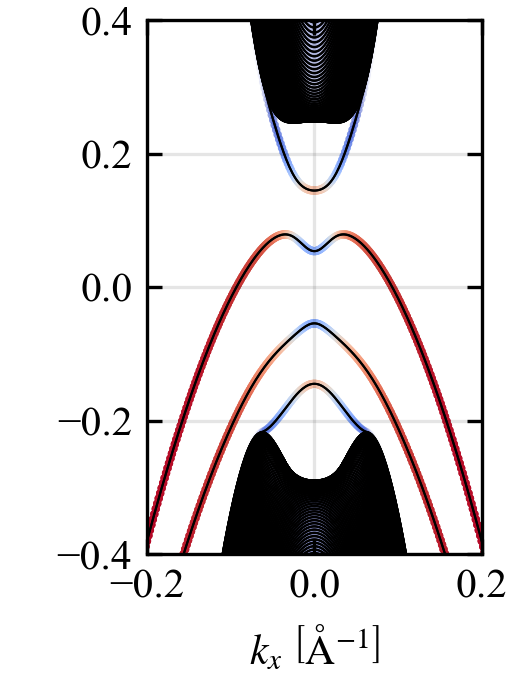}
		\label{fig:SG_Met_Band_t=0.1}
	\end{subfigure}
	\caption{\subref{fig:SurfaceGap} Variation of the width of the global surface gap by shifting the position of FMM band and changing the strength of exchange interaction in thin-film FMM. We took the adjustable parameters in Eqs. \eqref{eq:HamiMat_FMM}, \eqref{eq:HamiMat_FMM_TB} and \eqref{eq:Hami_IH} as $t=-1.0~\left [ \si{eV} \right ]$, $\bsym{M}=M_{z}\bsym{e}_{z}$, $\lambda_{\mathrm{R}}=0.2~\left [ \si{eV} \right ]$ and $t_{\mathrm{IH}}=0.1~\left [ \si{eV} \right ]$. The label of horizontal axis $\epsilon_{\mathrm{FMM},\downarrow}(\bar{\Gamma})$ represents the value of upper FMM band at the $k_{x}=k_{y}=0$, i.e., $\epsilon_{\mathrm{FMM},\downarrow}(\bar{\Gamma})=\epsilon_{0,\mathrm{FM}}-3t+JM_{z}$ from Eqs. \eqref{eq:HamiMat_FMM} and \eqref{eq:HamiMat_FMM_TB}. \subref{fig:SG_Ins_Band_t=0.1}, \subref{fig:SG_Met_Band_t=0.1} Band structure in case of setting the FMM band \subref{fig:SG_Ins_Band_t=0.1} far from or \subref{fig:SG_Met_Band_t=0.1} near to the original Dirac point localized on isolated 3DTI surface. We took the strength of exchange interaction as $JM_{z}=0.1~\left [ \si{eV} \right ]$ and the onsite energy in the thin-film FMM layers as \subref{fig:SG_Ins_Band_t=0.1} $\epsilon_{0,\mathrm{FMM}}=-3.3~\left [ \si{eV} \right ]$ or \subref{fig:SG_Met_Band_t=0.1} $\epsilon_{0,\mathrm{FMM}}=-3~\left [ \si{eV} \right ]$. The vertical axes of these figures have their scales in common.}
	\label{fig:Comparison}
\end{figure}
Finally, we discuss the requirement to induce a wide QAH gap in the 3DTI/FMM bilayer structure. For this purpose, we compare the width of the surface gap by shifting the position of FMM band within the energy range of the bulk gap of the 3DTI and changing the exchange-interaction strength in the FMM layer independently. Fig. \bothref{fig:Comparison}{fig:SurfaceGap} shows the width of the surface gap by changing the positions of the FMM band and different values of the exchange-interaction strength. The label $\epsilon_{\mathrm{FMM},\downarrow}(\bar{\Gamma})$ of the horizontal axis in the figure represents the energy value of upper FMM band at the $k_{x}=k_{y}=0$ point in SBZ; $\epsilon_{\mathrm{FMM},\downarrow}(\bar{\Gamma})=\epsilon_{0,\mathrm{FM}}-3t+JM_{z}$. Here we fixed the magnetization vector in the thin-film FMM as $\bsym{M}=M_{z}\bsym{e}_{z}$. We determined each width of the surface gap as an energy range where the anomalous Hall conductivity is constant and quantized as a unit of $e^{2}/h$ and the density of states vanishes. From Fig. \bothref{fig:Comparison}{fig:SurfaceGap}, we can anticipate that a sufficiently wide QAH gap of $80$ - $100~\left [ \si{meV} \right ]$ may appear in the 3DTI/thin-film FMM bilayer structure because the strength of exchange interaction can reach about $10^{0}~\left [ \si{eV} \right ]$ as order. On the other hand, when the original FMM band and the conduction band of bulk 3DTI intersect, which corresponds to the interval of $\epsilon_{\mathrm{FMM},\downarrow}(\bar{\Gamma}) \gtrsim 0.3~\left [ \si{eV} \right ]$ in Fig. \bothref{fig:Comparison}{fig:SurfaceGap}, the surface gap sharply vanishes because the FMM band completely covers the bulk gap of 3DTI after the two materials are attached.

Fig. \bothref{fig:Comparison}{fig:SurfaceGap} also shows that the width of the surface gap becomes maximum when the original FMM band lies on the original Dirac point which is initially localized at the 3DTI surface. We qualitatively discuss the mechanism for such variation of the surface gap in Fig. \bothref{fig:Comparison}{fig:SurfaceGap} by considering the band structure and the weight distribution for two types of FMM band. One is the case where the original FMM band lies far from the Dirac point on 3DTI surface at the $\bar{\Gamma}$ point, and the other is the one where the FMM band and the Dirac point intersect initially. Fig. \bothref{fig:Comparison}{fig:SG_Ins_Band_t=0.1} and \bothref{fig:Comparison}{fig:SG_Met_Band_t=0.1} show the obtained band structure with black lines and the weight distribution with color points on the lines for the two setting types of the FMM band. Fig. \bothref{fig:Comparison}{fig:SG_Ins_Band_t=0.1} clearly shows that the eigenstates around the gapped-out Dirac cone is still strongly localized at the top surface of 3DTI when the FMM band is initially separated from the original Dirac point. On the other hand, Fig. \bothref{fig:Comparison}{fig:SG_Met_Band_t=0.1} shows that when the FMM resides close to the Dirac point, the prominent migration of the DSS occurs as we have shown in Sec. \ref{sec:TPE}, which leads to the wide surface gap. This result indicates that intense exchange interaction on the DSS requires the migration of DSS realized by the hybridization between the thin-film FMM band and the DSS only around the original Dirac point. Once the DSS migrates into the FMM layer by such hybridization through the the interface coupling, a wide QAH gap opens because the exchange interaction can strongly acts on the DSS in the thin-film FMM layer.


\section{Conclusion} \label{sec:Conclusion}
In this work, we have suggested possibility of the occurrence of the QAH effect in the 3DTI/thin-film FMM bilayer structure and revealed its mechanism. When the surface dispersion of 3DTI and the band of thin-film FMM initially intersects near to the original Dirac point, the hybridization of the degenerated states cause the shift of the DSS into the FMM layer. This proximity effect allows the exchange interaction in the FMM to strongly act on the migrated DSS, which brings the appearance of the wide gap at the migrated Dirac dispersion. When the chemical potential lies in the surface gap, the anomalous Hall conductivity can be quantized due to the topological character of the non-trivial helical DSS. The realization of wide surface gap in the 3DTI/thin-film FMM bilayer structure leads to the easy observation of the QAH effect, e.g., at the room temperature.

Our computed results demonstrate that the width of the surface gap reach about $90~\left [ \si{meV} \right ]$ by setting practical material-dependent parameters which are expected from the result of \textit{ab initio} calculation in the 3DTI and the FMM of our model. Because the 3DTI/thin-film FMM bilayer structure can influence strong magnetic coupling in FMM to the DSS of 3DTI without any impurities, it can experimentally procure the wide QAH gap by solving not only the issue of disorder in the magnetically-doped TI but also the torment of suppressed magnetic penetration in 3DTI/ferromagnetic insulator heterostructures. Due to this advantage, we conclude that the 3DTI/thin-film FMM bilayer structure is one of the promising candidates to experimentally realize the QAH insulator within a wide range of temperature, which pave a way to further researches of other exotic topological phenomena and application of the less-dissipative edge conduction.

\begin{acknowledgments}
We are grateful to Koji Kobayashi for many helpful discussions. This work was supported by JSPS KAKENHI Grants No. JP20H01830 and No. 19H00650, JST CREST Grant No. JPMJCR18T2, and JST-Mirai Program Grant No. JPMJMI20A1.
\end{acknowledgments}

\bibliographystyle{apsrev4-2}
\bibliography{./library}

\end{document}